\documentclass[%
 aip,
 amsmath,amssymb,
 reprint,%
]{revtex4-1}

\usepackage{graphicx}% Include figure files
\usepackage{dcolumn}% Align table columns on decimal point
\usepackage{bm}% bold math
\usepackage[utf8]{inputenc}
\usepackage[T1]{fontenc}
\usepackage{mathptmx}
\usepackage{xcolor}
\usepackage{braket}
\usepackage{array}

\begin{document}

\preprint{AIP/123-QED}

\title[Squeezed Light Induced Two-photon Absorption Fluorescence of Fluorescein Biomarkers]{Squeezed Light Induced Two-photon Absorption Fluorescence of Fluorescein Biomarkers}
% Force line breaks with \\

\author{Tian Li}
%\altaffiliation[Also at ]{Physics Department, XYZ University.} %Lines break automatically or can be forced with \\
\email{tian.li@tamu.edu.}
\affiliation{ 
Department of Biological and Agricultural Engineering, Texas A\&M University, College Station, TX 77843, USA%\\This line break forced with \textbackslash\textbackslash
}%
\affiliation{%
Institute for Quantum Science and Engineering,  Texas A\&M University, College Station, TX 77843, USA%\\This line break forced% with \\
}%

\author{Fu Li}%
%\email{Second.Author@institution.edu.}
\affiliation{%
Institute for Quantum Science and Engineering,  Texas A\&M University, College Station, TX 77843, USA%\\This line break forced% with \\
}%
\affiliation{ 
Department of Physics and Astronomy, Texas A\&M University, College Station, TX 77843, USA%\\This line break forced with \textbackslash\textbackslash
}%

\author{Charles Altuzarra}
% \homepage{http://www.Second.institution.edu/~Charlie.Author.}
\affiliation{%
Institute for Quantum Science and Engineering,  Texas A\&M University, College Station, TX 77843, USA%\\This line break forced% with \\
}%
\affiliation{%
School of Physics and Astronomy, University of Glasgow, Glasgow G12 8QQ, U.K.%\\This line break forced% with \\
}%

\author{Anton Classen}
% \homepage{http://www.Second.institution.edu/~Charlie.Author.}
\affiliation{ 
Department of Biological and Agricultural Engineering, Texas A\&M University, College Station, TX 77843, USA%\\This line break forced with \textbackslash\textbackslash
}%
\affiliation{%
Institute for Quantum Science and Engineering,  Texas A\&M University, College Station, TX 77843, USA%\\This line break forced% with \\
}%

\author{Girish S. Agarwal}
% \homepage{http://www.Second.institution.edu/~Charlie.Author.}
\affiliation{ 
Department of Biological and Agricultural Engineering, Texas A\&M University, College Station, TX 77843, USA%\\This line break forced with \textbackslash\textbackslash
}%
\affiliation{%
Institute for Quantum Science and Engineering,  Texas A\&M University, College Station, TX 77843, USA%\\This line break forced% with \\
}%
\affiliation{ 
Department of Physics and Astronomy, Texas A\&M University, College Station, TX 77843, USA%\\This line break forced with \textbackslash\textbackslash
}%

\date{\today}% It is always \today, today,

\begin{abstract}
%\section*{Abstract}
Two-photon absorption (TPA) fluorescence of biomarkers has been decisive in advancing the fields of biosensing and deep-tissue \textit{in vivo} imaging of live specimens. %, possibly briefly mention other field, such TPA spectrsocopy or TPA in solids, etc]}. 
However, due to the extremely small TPA cross section and the quadratic dependence on the input photon flux, extremely high peak-intensity pulsed lasers are imperative, which can result in significant photo- and thermal damage. %(to biological specimen). 
Previous works on entangled TPA (ETPA) with spontaneous parametric down-conversion (SPDC) light sources found a linear dependence on the input photon-pair flux, but are limited by low optical powers, along with a very broad spectrum. We report that by using a high-flux %\textcolor[rgb]{0,0,1}{two-mode} 
squeezed light source for TPA, a fluorescence enhancement of $\sim47$  is achieved in fluorescein biomarkers as compared to classical TPA. %(with CW excitation). 
Moreover, a polynomial behavior of the TPA rate is observed in the DCM laser dye. 
\end{abstract}

\maketitle

%\section{Introduction}

Two-photon absorption (TPA) microscopy (with near-infrared illumination) is the method of choice for \textit{in vivo} imaging of tissues down to millimeter depths~\cite{Helmchen:2005by}. It bears several advantages including intrinsic high 3-D resolution due to significant TPA occuring only in close vicinity to the focal volume, reduced out-of focus bleaching, highly reduced autofluorescence, and the capability of nearly aberration-free deep-tissue focusing along with reduced absorption~\cite{CLSM, Xu10763,Zipfel:2003dq,Drobizhev:2011ai}.
%such as reduced out-of-focus photobleaching, less autofluorescence, deeper tissue penetration and intrinsically high 3-dimensional resolution. 
Unfortunately, classical TPA is an extremely inefficient process with absorption cross sections $\delta_r$ on the order of  $10^{-48}~\text{cm}^4\cdot\text{s}/\text{photon}$~\cite{Upton:2013fk}. % and scattering in biological tissues is very prominent. 
Therefore, TPA sensing and imaging generally requires the use of high-intensity % light sources, most frequently pulsed femtosecond 
pulsed lasers, to insure the near-simultaneous presence of two photons to induce the process~\cite{Xu:96,Albota:98}. However, since the excitation pulse peak power %in each pulse 
is typically $10^5$ times the average power, samples are prone to endure significant photo- and thermal damage~\cite{TAYLOR20161,doi:10.1111/php.12572}.
%According to Xu \textit{et al.}~\cite{Xu:96}, the alternative of using CW as opposed to pulsed sources is problematic because a factor of $10^2$ to $10^3$ times more average power is required to yield the same amount of fluorescence signal as for pulsed excitations. However, since for a pulsed excitation the power \textit{in each pulse} is typically $10^{5}$ (in the order of $1/f\tau$, where $f$ is the pulse repetition rate and $\tau$ is the excitation pulse width) times more than its average power, the energy in each pulse is still $10^2$ to $10^3$ times more than that of a CW excitation to acquire the same amount of fluorescence signal. \textcolor{red}{Thus using a CW excitation can therefore significantly reduce the risk of photodamaging, such as phototoxicity and photobleaching of the sample~\cite{TAYLOR20161, Chen1257998}, but unfortunately this option yields very little fluorescence}. 

In parallel, using the unique quantum energy-time entanglement characteristics of photon pairs generated by spontaneous parametric down-conversion (SPDC), \textcolor{black}{the entangled TPA (ETPA) rate can be vastly enhanced~\cite{Villabona-Monsalve:2018fy,Schlawin:2018ve,Upton:2013fk,Varnavski:2017rp,Villabona-Monsalve:2017fq}, with the absorption cross section $\sigma_e$ for ETPA in the range of $10^{-18} - 10^{-22}~\text{cm}^2$.} Most notably, the linear dependence of ETPA on the input photon-pair flux, which was first predicted by Gea-Banacloche~\cite{PhysRevLett.62.1603} and Javanainen and Gould~\cite{PhysRevA.41.5088}, was also verified experimentally   
%ETPA shows a linear dependence on the input photon-pair flux
~\cite{PhysRevLett.75.3426,PhysRevLett.78.1679,Villabona-Monsalve:2017fq,Upton:2013fk,Varnavski:2017rp,Villabona-Monsalve:2018fy}. It is a major advantage over the quadratic dependence of classical TPA as the need for high intensity excitation becomes obsolete. However, \textcolor{black}{most} current ETPA implementations with biological specimen are limited by a low flux of $\sim10^7$ photon pairs$/\text{s}$~\cite{Villabona-Monsalve:2018fy,Upton:2013fk,Varnavski:2017rp,Villabona-Monsalve:2017fq}, equivalent to only $\sim10$~pW for near-infrared wavelengths. %, which is unviable for bioimaging and biosensing. 
This is mostly due to \textcolor{black}{loss of correlation and difficulty of tuning biphoton wavelength in the nonlinear crystals.} %the low conversion efficiencies of the nonlinear crystals used to generate the entangled photon pairs. %It has also been shown that ETPA is unique in that it follows a linear behavior, unlike the quadratic behavior specific to classical TPA. 
\textcolor{black}{It is also worth mentioning that a much more efficient photon pair flux generation has been demonstrated by Jechow~\textit{et al.} using a type-0 quasi-phase-matched periodically-poled-lithium-niobate waveguide crystal~\cite{Jechow:s}. Their photon-pair flux can be as high as $\sim10^{11}$ photon pairs/s.}

%On the other hand, by using quantum-correlated photon pairs, the two-photon absorption rate can be vastly enhanced since the absorption process depends linearly rather than quadratically on the input photon-flux density~\cite{PhysRevLett.75.3426,PhysRevLett.78.1679,Ceskoslovensky.47.3.1997}. Proof-of-principle experiments have already demonstrated the linear behavior with energy-time entangled photon pairs produced from spontaneous parametric down-conversion (SPDC) in a Barium Borate (BBO) crystal~\cite{Villabona-Monsalve:2017fq,Upton:2013fk,Varnavski:2017rp,Villabona-Monsalve:2017fq}. Some fundamental experimental limitations include very low input entangled photon pair flux in the order of~$10^7$ photons/$s$, and very wide excitation linewidths. %Moreover, none of the molecules would be particularly useful as fluorescent biomarkers in the two-photon absorption microscopy~\cite{CLSM}. 

\begin{figure}
    \begin{center}
    \includegraphics[width=1\columnwidth]{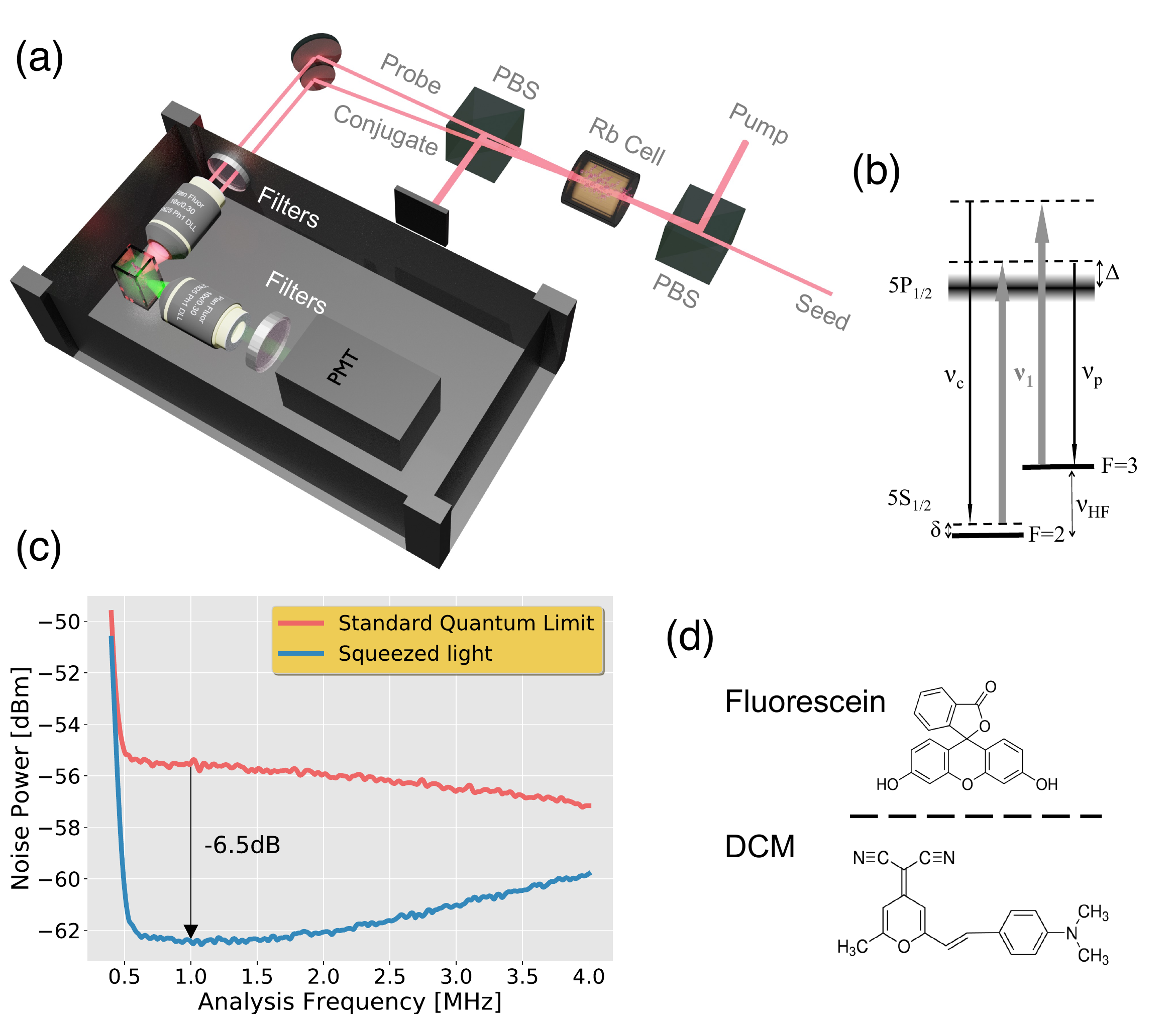}
    \caption{(a) Squeezed-light TPA setup in which a seeded $^{85}$Rb cell produces strong quantum-correlated twin beams via FWM. The twin beams are focused onto the sample with a 10$\times$ objective. Fluorescence is collected at an angle of $90^\circ$ with a second 10$\times$ objective and fed into a photomultiplier tube (PMT). Two short-pass filters in front of the PMT exclude stray pump photons. The setup is enclosed in a light-proof box. (b) Level structure of the D1 transition of $^{85}$Rb atoms. The optical transitions are arranged in a double-$\Lambda$ configuration, where $\nu_p$, $\nu_c$ and $\nu_1$ stand for probe, conjugate and pump frequencies, respectively, fulfilling $\nu_p$ +  $\nu_c$ =  $2\nu_1$. The width of the excited state in the level diagram represents the Doppler broadened line. $\Delta$ is the one-photon detuning, $\delta$ is the two-photon detuning, and $\nu_{\text{HF}}$ is the hyperfine splitting in the electronic ground state of $^{85}$Rb. (c) Measured intensity-difference noise power spectrum for the squeezed twin beams (blue line) and for the standard quantum limit (red line), obtained with a radio frequency spectrum analyzer (with a resolution and video bandwidth of 300~kHz and 100~Hz, respectively). A squeezing of 6.5~dB is achieved. (d) Molecular structures of fluorescein and DCM.} 
		\label{Setup}
    \end{center}
\end{figure}

In this work, we utilize a different quantum light source that is based on the four-wave mixing (FWM) process in an atomic $^{85}$Rb vapor cell~\cite{Dowran:18,PhysRevA.95.023803,Hudelist:2014db,Anderson:17,Li:17,Pooser:15,Clark:2014vf}. The setup and the respective atomic level structure are shown in Fig.~\ref{Setup}(a) and (b). The medium possesses a large third-order electric susceptibility $\chi^{(3)}$ and is pumped by a strong narrow-band continuous-wave (CW) laser at frequency $\nu_1$ ($\lambda = 795$~nm) with a typical linewidth $\Delta \nu_1 \sim 100$~kHz. Applying an additional coherent CW seed beam %of $\sim 1$~mW optical power, 
at frequency $\nu_p = \nu_1 - (\nu_{HF}+\delta)$, where $\nu_{HF}$ and $\delta$ are the hyperfine splitting in the electronic ground state of $^{85}$Rb and the two-photon detuning respectively in Fig.~\ref{Setup}(b) (see the Supplementary Material for further experimental details), two pump photons are converted into a pair of twin photons, namely `probe $\nu_p$' and `conjugate $\nu_c$' photons, adhering to the energy conservation $2 \nu_1 = \nu_p + \nu_c$ (see the level structure in Fig.~\ref{Setup}(b)). The resulting ``twin beams'' are strongly quantum-correlated and are also referred to as (seeded) two-mode squeezed light~\cite{PhysRevA.78.043816}. Major advantages are narrow-band probe and conjugate beams ($\sim 20$~MHz)~\cite{Clark:2014vf,Glasser2012a} along with a freely adjustable photon-pair flux between $10^{13}$ to~$10^{16}$ photons/s~\cite{Dowran:18,PhysRevA.95.023803,Anderson:17,Li:17,Pooser:15,Clark:2014vf}, which is a few orders of magnitude higher than for SPDC. \textcolor{black}{Also, since FWM in atomic vapors is an nonlinear parametric process based on ground-state coherences~\cite{McCormick:s}, in which the main advantage arises from small two-photon detunings from real states whereas in nonlinear crystals there is no real state to which the signal or idler photon is close, the generation of quantum correlations with FWM in atomic vapors can be  therefore very efficient.} As can be seen from Fig.~\ref{Setup}(c), the twin beams exhibit a intensity-difference squeezing of 6.5~dB, which is indicative of strong quantum correlations~\cite{PhysRevA.78.043816} (see the Supplementary Material for further details of the squeezing measurement). % \textcolor[rgb]{0,0,1}{(meaning for induced TPA?)}
% \textcolor[rgb]{0,0,1}{[Needs to be mentioned later and elaborated more: we further report that the relationship between input power and fluorescence intensity with squeezed light follows a linear behavior for fluorescin and a nonlinear one for DCM]}

% and when appropriately chosen laser light `seeds' the medium, and squeezed light is produced (see Fig.~\ref{Setup}(a) for a typical squeezing spectrum and experimental details can be found in Supplementary Material). Practically, FWM has proven to be an excellent platform for quantum sensing applications because of its great potential for generating squeezed light and entanglement, and for its experimental simplicity and integrability~\cite{Dowran:18,%Li:17,
% PhysRevA.95.023803,Anderson:17,Pooser:15,Clark:2014vf}.  As can be seen from Fig.~\ref{Setup}(a), the entangled photon pair flux of the source exhibits a strong intensity-difference squeezing of -6.5~dB, which is indicative of  strong quantum correlations between the twin beams. Due to the much higher production of photon pairs in the squeezed light source and differently from the entirety of ETPA experiments governed by a low photon pair flux, we further report that the relationship between input power and fluorescence intensity with squeezed light follows a nonlinear behavior.

%\section{Experiment}

\begin{figure}
    \begin{center}
    \includegraphics[width=0.86\columnwidth]{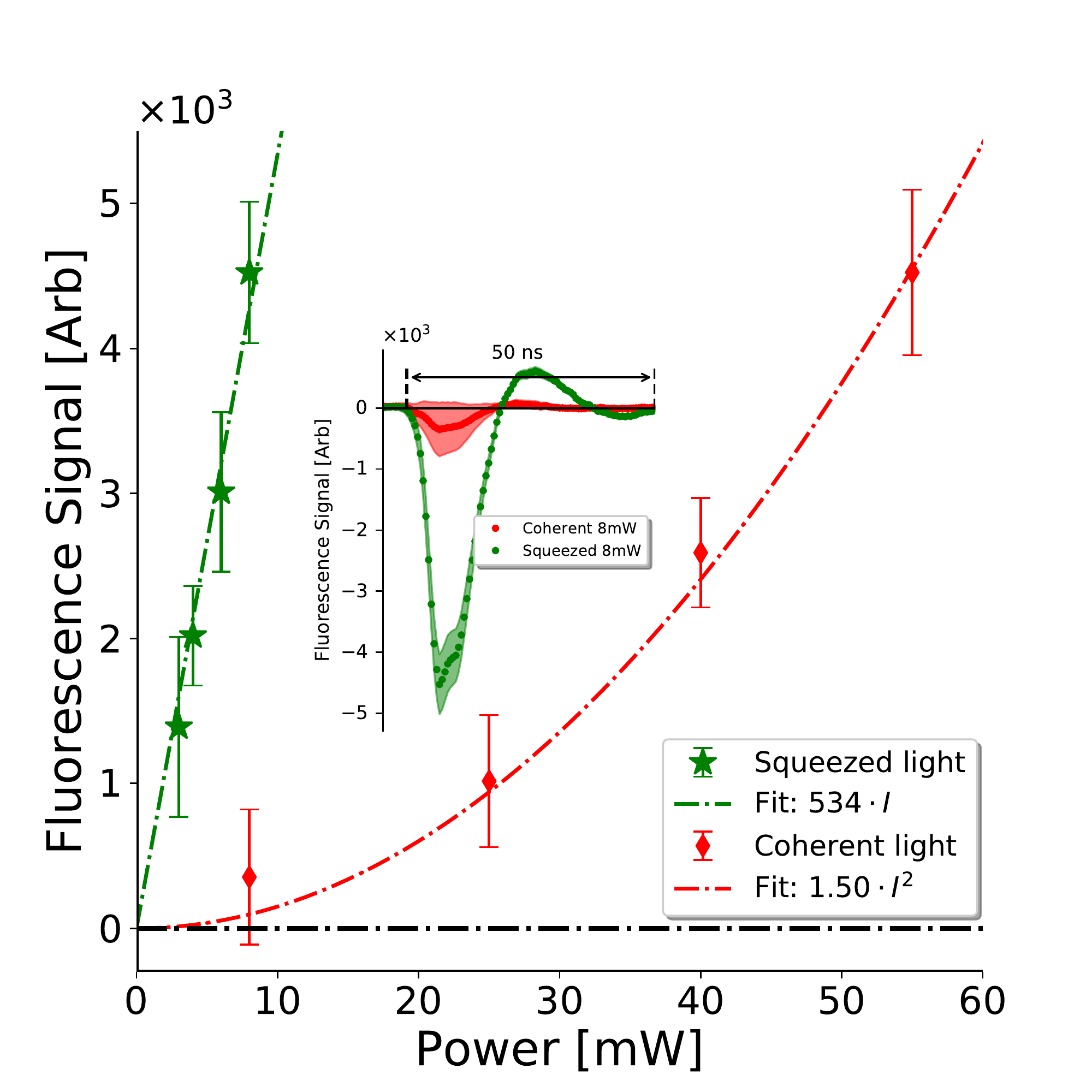}
    \caption{
        \label{Fluorescein_a}
    %
    % (Color online)
    %
   Fluorescence rates versus excitation power. The red diamonds and the red (dash-dotted) line show the measured values for the coherent excitation and the respective quadratic fit. The green stars are the rates for SL-TPA induced by the twin-beam excitations, and the green (dash-dotted) line is the respective linear fit. Error bars denoting one standard deviation. %, and enhancement of a factor of $\sim$~50 with squeezed light at 8~mW. 
	%The fluorescence signal from 4~mW of squeezed light excitation is also shown. 
Inset: raw voltage output from the PMT for coherent light (red) and squeezed light (green) excitations of 8~mW optical power. The shaded area for each curve represents one standard deviation. 
}
    \end{center}
\end{figure}

In this study we analyze and compare classical TPA and squeezed-light induced TPA (SL-TPA) fluoresecence rates in fluorescein and DCM (see the Supplementary Material for samples preparation). Fluorescein is one of the most frequently used biomarkers for bioimaging and biosensing~\cite{Cells.2.591.2013}. % \textcolor[rgb]{0,0,1}{(rather mention TPA microscopy?)}. 
Its small size is very convenient for \textit{in vivo} imaging applications, although its relatively small classical TPA cross section generates low amounts of fluorescence~\cite{Xu:96,Albota:98}. The SL-TPA setup is depicted in Fig.~\ref{Setup}(a). A 10$\times$ objective (\textcolor{black}{Thorlabs RMS10X}) focuses the near-infrared excitation light onto a solution of fluorophores. Following TPA, fluorescence is collected by a second 10$\times$ objective (\textcolor{black}{Thorlabs RMS10X}) at an angle of $90^\circ$ and guided to a photo-multiplier tube (PMT) (\textcolor{black}{Thorlabs PMTSS in conjugation with a  PMT transimpedance amplifier  Thorlabs TIA60}). Two optical low-pass filters (\textcolor{black}{Thorlabs FESH0750}) exclude stray pump photons. The measured PMT voltage outputs (see inset in Fig.~\ref{Fluorescein_a}) are converted into fluorescence rates of arbitrary units (\textcolor{black}{since the PMT's response to an input photon is an inverse voltage pulse, adding all the inverse voltages in a given time window can therefore give us a quantity that is proportional to the input photon flux up to a conversion factor}, see the Supplementary Material for further data acquisition details). For classical TPA measurements only the coherent pump beam is focussed into the microscope objective, with the same focus spot size at the sample. Utilized powers for the twin beams were ranging from $30~\mu$W to the maximum of 8~mW. %, and thus comparable with the average optical power of the twin beams.
% In addition, all measurements conducted with squeezed light are compared with measurements conducted with a continuous-wave (CW) laser of the same intensity, which in this work the classical coherent light source. 

%\subsection{Fluorescein in water}

Measured classical TPA fluorescence rates for fluorescein, as a function of the input power, are shown by the red diamonds in Fig.~\ref{Fluorescein_a}, with error bars denoting one standard deviation. The observed quadratic power law relationship agrees well with the established literature for classical TPA, where the fluorescence signal is proportional to the square of excitation light intensity~\cite{PhysRev.175.1555}. Fluorescence rates induced by SL-TPA of 8~mW optical power (3.5~mW + 1.0~mW probe and seed; 3.5~mW conjugate) and 4~mW optical power (1.75~mW + 0.5~mW probe and seed; 1.75~mW conjugate) \textcolor{black}{together with 6~mW optical power (2.6~mW + 0.8~mW probe and seed; 2.6~mW conjugate) and 3~mW optical power (1.3~mW + 0.4~mW probe and seed; 1.3~mW conjugate) are depicted by the green stars (although the 3 mW and 6 mW data points were taken on a different day, the trend is similar).} Due to experimental constraints, 8~mW is the maximal power we are able to acquire for the squeezed light. 
%and its fluorescence is compared with that generated from 8~mW of coherent light. 
The measured coherent rates are fitted by the quadratic function $R(I) = I^2 \times 1.5~mW^{-2}$ (dash-dotted red line), which represents the benchmark of the true fluorescence rate as a function of the input power. It can be observed from the figure that the signal for 8~mW of coherent excitation deviates strongest from the fit. This fact can be attributed to background noise (e.g., electronic dark counts and spurious counts from stray ambient light) and the overall low signal to noise ratio (SNR) of the measurement (characterized by a standard deviation encompassing negative values). Following the fitted curve, the fluorescence rate for 8~mW coherent excitation is thus merely $9.6 \times 10^{1}$~[a.u.]. For SL-TPA of 8~mW excitation power the fluorescence rate is $\sim 4.46\times 10^{3}$~[a.u.]. This value corresponds to a striking $\sim 46.5$-fold enhancement over the coherent excitation. Vice versa, increasing the coherent excitation power seven-fold to $\sim 55$~mW, and thus the classical TPA rate by $\sim 47.3$-fold, the measured rates for 8~mW SL-TPA and 55~mW classical TPA match, thus confirming the previous statement. Subtracting the contribution from the 1~mW coherent seed beam power it can be argued that the measured SL-TPA enhancement is around $\sim 60$-fold over 7~mW coherent excitation. Note that the seed is uncorrelated to the quantum correlated photon pairs and that the 1~mW of coherent seed excitation itself will induce negligible classical TPA rates. 
%However, when fluorescein is excited with 8~mW of squeezed light, the fluorescence signal (represented by the green star) is enhanced by a factor of $\sim$~50 as compared to the true value of TPA fluorescence with 8~mW of coherent excitation. 
More importantly, the measured fluorescence rate for 4~mW SL-TPA is $\sim 2.02\times 10^{3}$~[a.u.]. Subtracting the respective optical power of the seed beam, we end up with the input flux ratio $7/3.5 = 2.00$ which matches the measured ratio $4.46/2.02 = 2.21$ well (within the calculated uncertainties). This is also true for the SL-TPA of 6~mW and 3~mW, which is indicative of the linear dependence on the input photon-pair flux that is expected for fluorescein in this regime. Quadratic terms thus do not seem to contribute here. %\textcolor[rgb]{0,0,1}{[What about subtraction of dark-counts and offset from the measured SL-TPA signals?]}
 %More importantly, matching the fluorescence signal from 8~mW of squeezed light with the CW laser requires at least a power of 55~mW, thus approximately 7 times more intensity. 

%\textcolor[rgb]{0,0,1}{[Possibly add a sentence on: ... and thus average power, which is highly detrimental since the acquired photodamage scales approximately with the square of the average illumination power [2016, Akhipov et al.]]}

\begin{figure}
    \begin{center}
    \includegraphics[width=0.85\columnwidth]{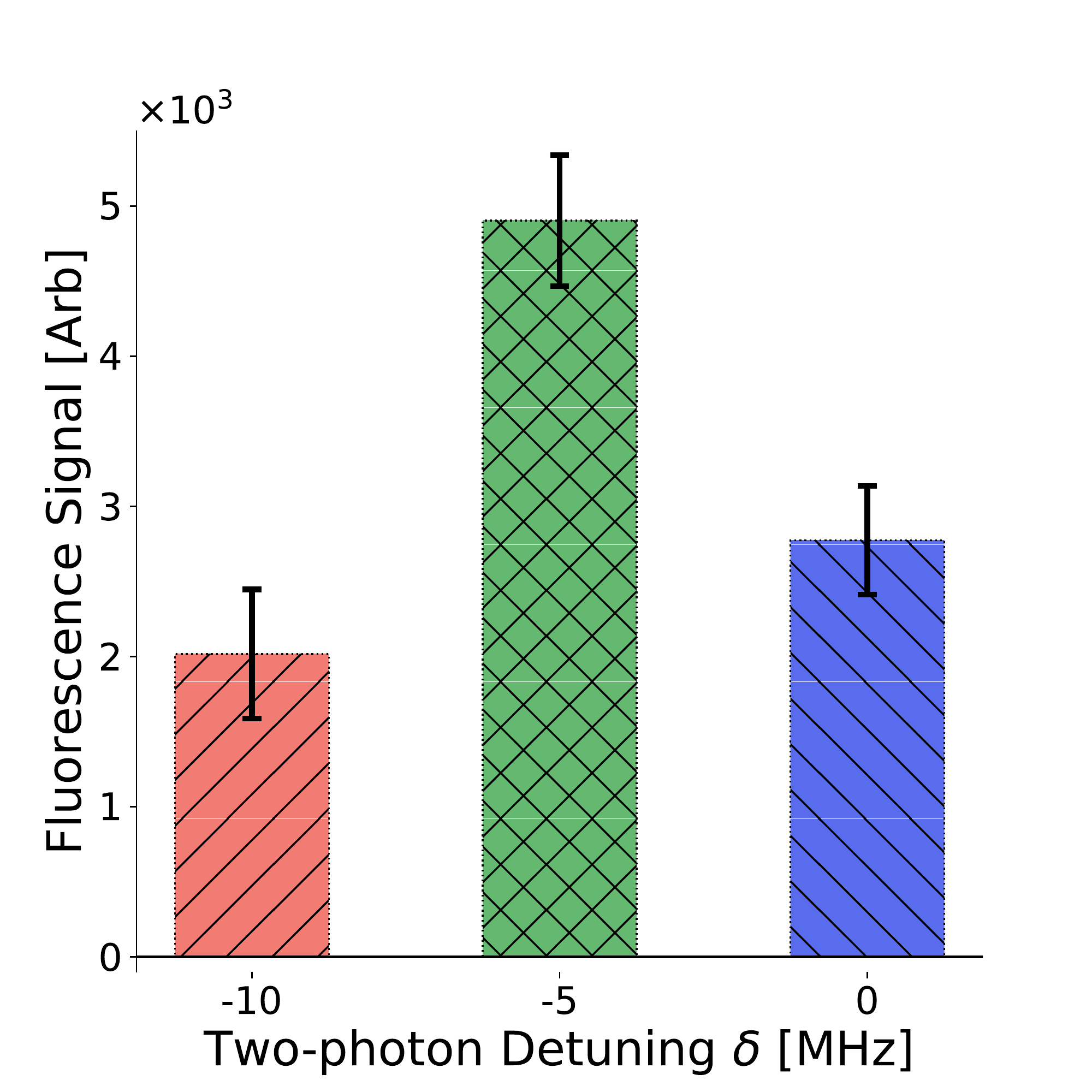}
    \caption{
        \label{Fluorescein_b}
    %
    % (Color online)
    %
    Fluorescence rates for 8~mW SL-TPA with three different two-photon detunings $\delta$, shown in the atomic level structure in Fig.~\ref{Setup}(b) as $\delta$. Red, green and blue bars are for $\delta$ = $-10$~MHz, $-5$~MHz and 0~MHz, respectively. This figure demonstrates degraded SL-TPA enhancements as a function of the relative arrival times of the entangled photon pairs. 
}
    \end{center}
\end{figure}

TPA is highly sensitive to the near-instantaneous arrival of two photons at the sample, in particular with respect to the virtual state lifetime of the intermediate states used for the electronic transition from the ground to the final state~\cite{PhysRevLett.78.1679,Varnavski:2017rp}. In ETPA this is quantified by the entanglement time $T_e$~\cite{Upton:2013fk,Varnavski:2017rp,PhysRevLett.123.023601}. Adjusting $T_e$ should change the measured SL-TPA enhancement. Hence, an investigation of the effect of relative temporal delay between the entangled photon pairs %on the enhancement of SL-TPA 
is conducted. The ETPA cross section $\sigma_e$ is inversely proportional to $T_e$ and thus the mean group velocity delay between the entangled photon pairs. %(i.e., the entanglement time $T_e$)~\cite{Ceskoslovensky.47.3.1997,Upton:2013fk,Varnavski:2017rp}, 
In the FWM process of the atomic $^{85}$Rb vapor, the group delay between the entangled pairs can be adjusted by changing the two-photon detuning $\delta$ of the double-$\Lambda$ configuration %in the atomic level structure, denoted in 
in Fig.~\ref{Setup}(b)%as $\delta$
~\cite{Glasser2012a}. The red, green and blue bars in Fig.~\ref{Fluorescein_b} show the fluorescence rates induced by 8~mW SL-TPA for the values $\delta$ = $-10$~MHz, $-5$~MHz and 0~MHz, respectively. For each $\delta$ value the same relative intensity-difference squeezing of 6.5~dB (see Fig.~\ref{Setup}(c)) is maintained, such that the results are not affected by different entanglement levels. %The red, green and blue bars represent the fluorescence signal for $\delta$ = -10~MHz, -5~MHz and 0~MHz, respectively. 
For $\delta = -5$~MHz a relatively small delay is achieved~\cite{Glasser2012a}. Degraded fluorescence rates for $\delta = -10$~MHz and $\delta = 0$~MHz confirm that the SL-TPA enhancement is degraded when the relative delay between the photon pairs is tuned away from its optimal value. Further experimental details on how to change the two-photon detuning $\delta$ can be found in the Supplementary Material.

%\subsection{DCM in DMSO}

In general, the ETPA rate $R_e$ as a function of the input photon-pair flux density $\phi$ is expected to follow the functional behavior $R_e(\phi) = \sigma_e \phi + \delta_r \phi^2$, where $\sigma_e$ and $\delta_r$ are the cross sections for ETPA and classical TPA respectively and are determined by the electronic level structure of the system undergoing TPA~\cite{Schlawin:2018ve,Harpham:2009zt,PhysRevLett.78.1679,PhysRevLett.75.3426}. Both values can be determined experimentally, or calculated theoretically via second-order perturbation theory for a sufficiently simple system~\cite{PhysRevLett.78.1679}. As previously established, the coincident arrival and absorption of an entangled photon pair leads to the linear dependence $R_e(\phi) = \sigma_e \phi$ provided $\phi$ is sufficiently small~\cite{Upton:2013fk,Varnavski:2017rp,Villabona-Monsalve:2017fq}. For sufficiently high photon-pair fluxes, TPA can be induced by uncorrelated photons from different pairs as well. The respective rate is equivalent to the classical TPA rate $\delta_r \phi^2$. Parity between both contributions is reached at the critical flux $\phi_c = \sigma_e / \delta_r$. Most previous ETPA experiments with biomarkers and low photon-pair fluxes observed only the linear dependence~\cite{Upton:2013fk,Varnavski:2017rp,Villabona-Monsalve:2017fq}.

  \begin{figure}[]
    \begin{center}
    \includegraphics[width =0.85\columnwidth]{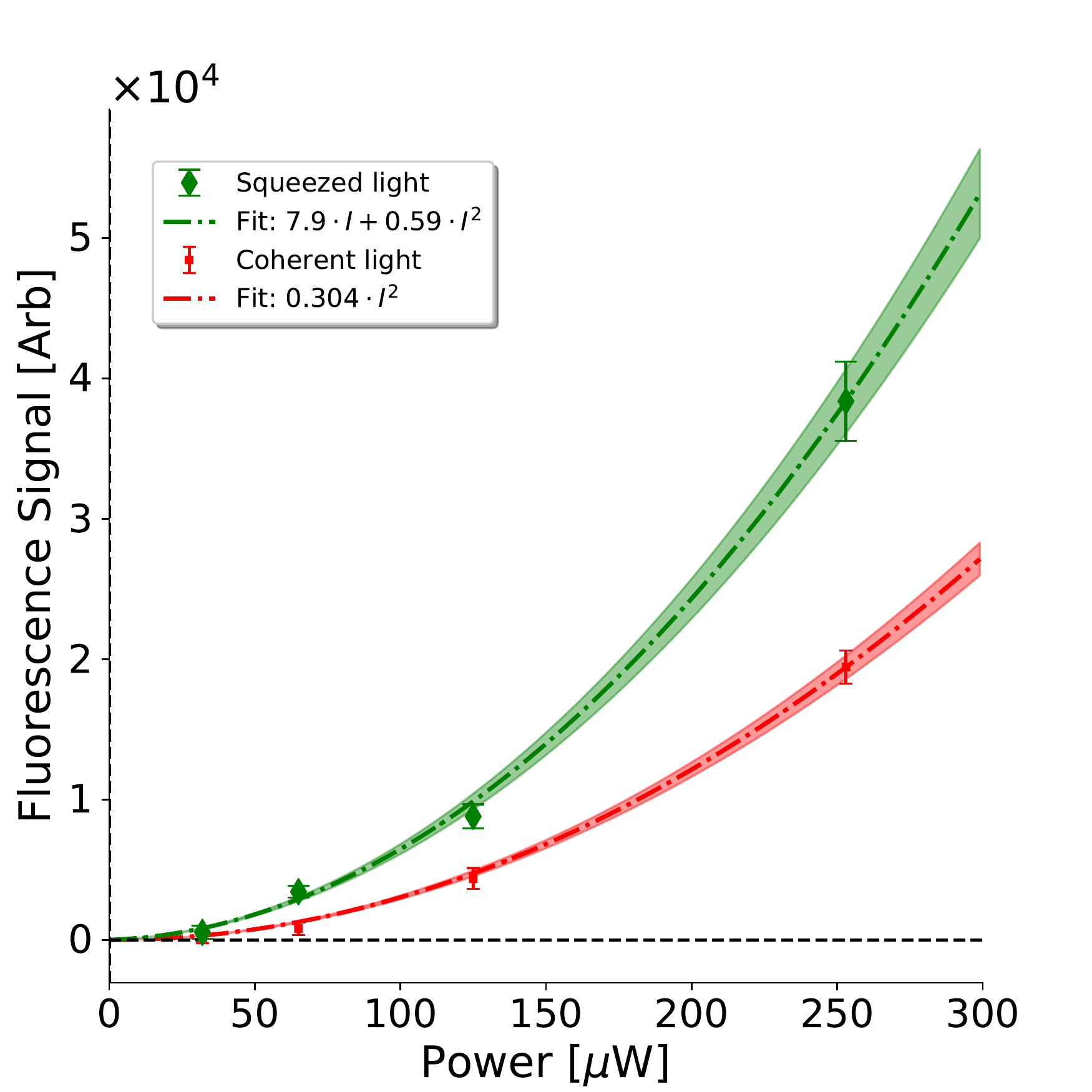}
    \caption{
        \label{DCM}
    %
    % (Color online)
    %
   Fluorescence signal of DCM versus excitation power of coherent light (red) and squeezed light (green). Coherent light fitting curve obeys a quadratic behavior, while squeezed light fitting curve shows a polynomial behavior, indicating a high input of entangled photon-pair flux.
        }
    \end{center}
\end{figure}

With means of investigating SL-TPA with high and freely adjustable optical powers, we investigated it's functional behavior in DCM laser dye (see the Supplementary Material for its preparation). %In addition, with means of validating the observation of TPA in our optical apparatus, a characterization of the nonlinearity is conducted by obtaining results with different coherent light excitation powers for DCM dyes (see Supplementary Material for its preparation). 
DCM dyes are known for strong TPA around 800~nm excitation wavelength, and along with a milli molar suspension, optical powers in the $\mu$W regime are sufficient to induce enough TPA fluorescence~\cite{Jechow:2013kx,Scheul:11}. The measured coherent TPA rates shown as red squares in Fig.~\ref{DCM}, agree well with a quadratic behavior, represented by the fit function $R(I) = I^2 \times 0.304~\mu\text{W}^{-2}$.
%Logically, like in the case of fluorescein, this investigation should yield a quadratic coherent excitation power to fluorescence signal relationship as well. Indeed, as can be observed from Fig.~\ref{DCM} with red squares, the fit is undeniably quadratic. 
For SL-TPA in DCM, on the other hand, we observed a non-linear behavior of the functional form $R(I) = I \times 7.9~\mu\text{W}^{-1} + I^2 \times 0.59~\mu\text{W}^{-2}$. %More importantly though, when the DCM dyes are excited with different powers of squeezed light, the excitation power to fluorescence signal relationship has a nonlinear behavior. 
%Indeed, ETPA can be accompanied by nonentangled photons or random classical two-photon absorption effects~\cite{Javanainen:1990gf}. 
%Therefore, the overall two-photon absorption rate, $R_e$, can be expressed~\cite{Harpham:2009zt,PhysRevLett.78.1679,PhysRevLett.75.3426,Ceskoslovensky.47.3.1997} as the summation of the linear ETPA rate and the quadratic classical TPA rate, $R_e = \sigma_e \phi + \delta_r \phi^2$, where $\sigma_e$ is the entangled two-photon absorption cross section, $\delta_r$ is the classical two-photon absorption cross section, and $\phi$ is the input photon flux density of photon pairs. 
%When the input photon flux is low, the linear term dominates~\cite{Upton:2013fk,Varnavski:2017rp,Villabona-Monsalve:2017fq}, while both contributions are significant when the input flux is high.  
The polynomial behavior of SL-TPA in Fig.~\ref{DCM} implies that the photon-pair flux is already high enough to observe both linear and quadratic contributions. In fact, given the fit values, parity is already reached at $I_c = 7.9 ~\mu\text{W}^{-1}/0.304 ~\mu\text{W}^{-2} = 26.0~\mu\text{W}$ for the DCM solution.

It is worthy to point out that the DCM laser dye solution requires much lower excitation powers than the fluorescein solution to produce appreciable TPA fluorescence rates, most probably due to a larger classical TPA cross section $\delta_r$. In Fig.~\ref{DCM}, the DCM signal at 130~$\mu$W coherent excitation ($\sim 0.45 \times 10^4$~a.u.) roughly equals the fluorescein signal at 55~mW coherent excitation ($\sim 4.43 \times 10^3$~a.u.). Taking into account the concentration of the two solutions (see the Supplementary Material for details of samples preparation)~\cite{paulRE}, we estimate the classical TPA cross section $\delta_r$ of DCM is roughly 1800 times larger than that of fluorescein. However, the ETPA cross section $\sigma_e$ of DCM is actually smaller than that of fluorescein, as demonstrated by the relatively small SL-TPA enhancements. %Extrapolation of the fit curves in Fig.~\ref{Fluorescein_a} would yield $I_c = 562~\text{mW}^{-1} / 1.5 ~\text{mW}^{-2} = 3.75 \times 10^{5}~\mu\text{W}$, which is $\sim 1.4 \times 10^{4}$ higher than for DCM. 
The difference can be attributed to different electronic level structures of these two organic molecules~\cite{Upton:2013fk}.

%As can be seen from Figs.~\ref{Fluorescein}(a) and~\ref{DCM}, the fluorescence signal induced by 130~$\mu$W of coherent light from DCM is greater than that induced by 8~mW of coherent light from fluorescein. 
%However, the entangled two-photon absorption cross section $\sigma_e$ of DCM is actually smaller than that of fluorescein, as demonstrated by the fluorescence enhancements. This inconsistency can be attributed to different electronic level structures of these two organic molecules~\cite{Upton:2013fk}. 

%\section{Conclusions}

In conclusion, this work investigates two-photon absorption fluorescence rates in fluorescein biomarkers and in DCM laser dye, induced by a coherent CW excitation light and by the bright two-mode squeezed light. For the coherent CW excitation both fluorophores show the well-expected quadratic dependence on the input photon flux. The experimental results for fluorescein with SL-TPA, however, demonstrate a linear dependence on the input optical power, along with a $\sim$~47-fold TPA fluorescence enhancement for 8~mW squeezed light compared to 8~mW coherent light. This can be attributed to the predominant occurrence of entangled two-photon absorption of quantum-correlated photon pairs. %playing the dominant role. 
%From extrapolation it can be concluded that parity between classical and entangled TPA contributions would only be reached at 375~mW optical power \textcolor[rgb]{1,0,0}{[also transform into flux via photon energy and the focus area]}.  
%that as compared to 8~mW of CW coherent light excitation, 
%8~mW of squeezed light achieves a $\sim$~60-fold TPA fluorescence enhancement. 
In addition, and differently from previous works using quantum states of light for TPA in fluorophores, we report that SL-TPA in DCM laser dye is governed by a polynomial behavior, which can be entirely attributed to its far greater entangled photon-pair flux, as compared to using SPDC sources. Thus, this work demonstrates that our FWM based bright two-mode squeezed light source can achieve ultra-low intensity TPA for biosensing and bioimaging, and thus bear the potential to open up unconventional avenues for \textit{in vivo} deep tissue studies of biological specimens via TPA.

%\section*{Supplementary Material}

See supplementary material for complete experimental details, squeezing measurements, data acquisition and samples preparation.

%\section*{Data Availability Statement}

The data that support the findings of this study are available from the corresponding author upon reasonable request.

This work is supported by the Air Force Office of Scientific Research 
(Award No. FA-9550-18-1-0141) and the Robert A. Welch Foundation (Award No. A-1943-20180324).

%\begin{acknowledgments}
We thank V. Yakovlev for discussions on spectroscopy with entangled photon pairs and  for the suggestion of Fluorescein biomarkers.  We also thank A. Sokolov and A. Zheltikov for informative comments. A. C. acknowledges support from the Alexander von Humboldt Foundation in the framework of a Feodor Lynen Research Fellowship.

\section*{References}
\bibliography{MyLibrary}

\end{document}